\documentclass[twocolumn]{aastex62}

\def\ltsima{$\; \buildrel < \over \sim\;$}
\def\ltsim{\lower.5ex\hbox{\ltsima}}
\def\gtsima{$\; \buildrel > \over\sim \;$}
\def\gtsim{\lower.5ex\hbox{\gtsima}}
\def\ms{$M_{\odot}$ }

\received{xxx}
\revised{xxx}
\accepted{2021 October 2}
\shorttitle{Two sites of r-process production}
\shortauthors{Tsujimoto}
\begin{document}

\title{Two Sites of r-Process Production Assessed on the Basis of the Age-tagged Abundances of Solar Twins}

\correspondingauthor{Takuji Tsujimoto}
\email{taku.tsujimoto@nao.ac.jp}

\author[0000-0002-9397-3658]{Takuji Tsujimoto}

\affiliation{National Astronomical Observatory of Japan, Mitaka, Tokyo 181-8588, Japan}



\begin{abstract}
Solar twins, i.e., stars that are nearly identical to the Sun, including their metallicities, in the solar vicinity show ages widely distributed from 0$-$10 Gyr. This fact matches the orbital history of solar twins in the new paradigm of galactic dynamics, in which  stars radially move on the disk when they encounter transient spiral arms. This finding suggests that older twins were born closer to the Galactic center and traveled a longer distance to reach their present location, according to the hypothesis that chemical enrichment occurs more quickly and that solar metallicity is attained on a shorter timescale with a decreasing Galactocentric distance ($R_{\rm GC}$). We show that abundance patterns covering a wide range of heavy elements for solar twins sharing similar ages are identical and that their variation among different age groups can be understood on the basis of the age-$R_{\rm GC}$ connection within the framework of Galactic chemical evolution. This study identifies the Galactic bulge as the birthplace of the oldest solar twins. Based on this scheme, we find that the relation between [$r$-process/Fe] and $R_{\rm GC}$ for the inner Galactic region is incompatible with the hypothesis of a sole site for $r$-process production, that is, neutron star mergers, whose delay time distribution could be approximated by the power-law form ($\propto t^{n}$). Alternatively, this relation suggests the presence of two distinct sites for $r$-process production:  short-lived massive stars, ending with specific core-collapse supernovae, and neutron star mergers that are heavily inclined to emerge with longer delay times, as represented by $n$=0$-$0.5.  
\end{abstract}



\section{Introduction}

One of the remarkable advancements in the astrophysics field in recent years is our understanding of the origin of $r$-process elements. After an intense debate on this issue for more than half a century, we eventually identified neutron star mergers (NSMs) as at least one site of $r$-process elements via the discovery of gravitational waves from NSM GW170817 and the subsequent detection of multiwavelength electromagnetic counterparts \citep[e.g.,][]{Smartt_17, Pian_17, Cowperthwaite_17}. Our recent interest has been directed toward the question of whether NSMs are the sole (major) site of the $r$-process \citep[e.g.,][]{Ji_19, Kobayashi_20, Cavallo_21}. This question arises from the following arguments. Theoretically, some specific core-collapse supernovae (CCSNe), such as magnetorotational SNe \citep[e.g.,][]{Winteler_12, Nishimura_15} or collapsars \citep[e.g.,][]{Siegel_19}, are capable of providing the physical conditions necessary to realize the $r$-process. The property of these CCSN candidates is a fast release of $r$-process products owing to a short lifetime of massive stars, in stark contrast to the long delay time, typically a few Gyrs, of NSMs \citep[e.g.,][]{Dominik_12}, which is exemplified by the GW170817 case \citep[6.5 Gyr;][11.2 Gyr; Fong et al.~2017]{Im_17}. In fact, the presence of a population exhibiting  such prompt $r$-process enrichment is implied by the stellar record of the abundances of these products in very metal-poor stars \citep[e.g.,][]{Roederer_13} as well as in Galactic disk stars \citep{Cote_19, Siegel_19}. On the other hand, the delay time distribution (DTD) of NSMs, ranging from a few 10 Myrs to a Hubble time, is not well constrained \citep[][and references therein]{Greggio_21}, allowing the possibility of NSMs being the only source of $r$-process elements. In the end, a full understanding of $r$-process sites is still lacking.  

Unlike $r$-process elements, the production sites of other heavy elements are well understood. For instance, $\alpha$-elements such as O and Mg are exclusively produced in CCSNe, while Fe-peak elements are produced in two sites, i.e., CCSNe and type Ia SNe (SNe Ia), whose DTDs are given by $\propto t_{\rm delay}^{-1}$ \citep{Maoz_14}. In addition, $s$-process elements represented by Ba are dominantly produced in asymptotic giant branch (AGB) stars in the late evolutionary stage of low- and intermediate-mass stars. Owing to the different release timescales among the multiple production sites for individual elements, the elemental abundance pattern from $\alpha$- to neutron-capture elements will change with the elapsed time, and its time variation will depend on the star formation rate in situ. Here, suppose that several detailed elemental abundance patterns that result from different star formation rates, i.e., different speeds of chemical enrichment, are observationally set up. Then, a comparison of the variation in $r$-process abundances generated by a slow to high enrichment speed with the variation in other well-understood elements would enable a precise assessment of the site of $r$-process element production associated with their release timescale.

Remarkably, this supposed setup is realized thanks to two recent advancements in the study of the dynamics and chemistry of Galactic stars. The improved understanding of Galactic dynamics suggests that stars radially move on the disk when they encounter transient spiral arms that are naturally generated during the process of disk formation \citep{Sellwood_02, Roskar_08, Grand_12, Baba_13}. This so-called radial migration of stars predicts that the birthplaces of a considerable number of disk stars are different from their current positions. According to this prediction, the stars in the solar vicinity represent the mixture of stars born at various Galactocentric distances ($R_{\rm GC}$) over the disk. The impact of this migration on the local chemistry is of significance since the stars born outside the solar vicinity follow chemical enrichment tracks that differ from the local tracks. This argument is based on our established knowledge that the chemical evolution of the disk differs in accordance with $R_{\rm GC}$, which is observationally evidenced by current radial abundance gradients showing higher metallicity at smaller $R_{\rm GC}$ values. This negative abundance gradient is theoretically understood in the context of galaxy formation, with the inner region forming faster and becoming more metal-rich than the outer region \citep[the so-called inside-out scenario;][]{Chiappini_01}. In other words, the rate at which the metallicity reaches a certain value (e.g., the solar metallicity) increases as $R_{\rm GC}$ decreases. 

Accordingly, for nearby stars sharing the same metallicity, we have two arguments: (i) the ages could vary among these stars as an outcome of radial migration, and (ii) older stars would have been born at smaller $R_{\rm GC}$ values. Therefore, the precise measurement of the age and elemental abundance of these local stars, if obtained, could provide elemental features as a function of $R_{\rm GC}$. At different $R_{\rm GC}$ values, chemical enrichment proceeds at a different speed. 

Locally, there exist stars that are nearly identical to the Sun, so-called solar twins, which exhibit stellar atmospheric characteristics quite similar to the solar values. Thanks to this close resemblance, the stellar ages of solar twins can be precisely determined, with a typical uncertainty of 0.4 Gyr, together with highly accurate (an error $<$ 0.01 dex) chemical abundances \citep{Bedell_18, Spina_18}. 
 Here, we show that the precise age-tagged abundances of solar twins offer new insight in the study of the $r$-process.
 
\section{Abundance patterns as a function of the solar twin's age}

We compare the chemical abundances of 28 elements, from carbon to dysprosium, with respect to Fe for 79 solar twins \citep{Bedell_18}. A quick overview indicates that stars with similar ages broadly share similar elemental abundance patterns, which is anticipated on the basis of individual [X/Fe] ratios as a function of the stellar age \citep{Bedell_18}. Then, we search for the representative patterns for each age group to understand how the patterns change with age via the following procedure. First, we separately collect the patterns for several age bins and choose a typical pattern for each group by eye. Then, we assign stars to each group by calculating deviations from a chosen one and select those below a fixed value of deviation. This procedure is repeated by altering age bins and/or the selection of typical patterns until the best classification of age-tagged abundance patterns is identified. 

\begin{figure}[t]
%	\vspace{-4.9cm}
    \vspace{0.3cm}
    \hspace{0.3cm}
%    \hspace{5.2cm}
	\includegraphics[width=0.91\columnwidth]{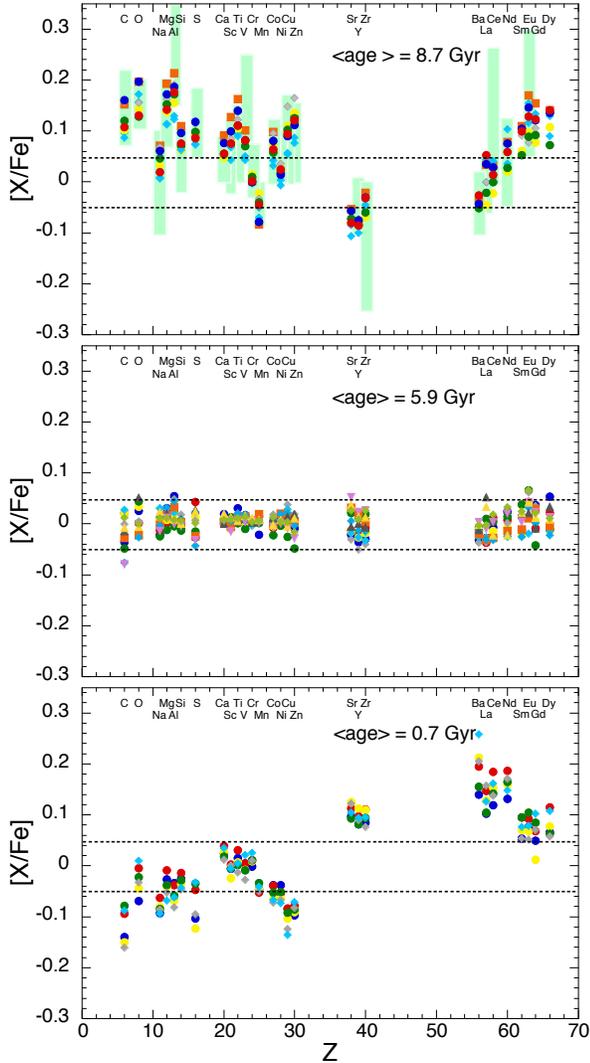}
%	\vspace{5.cm}
\caption{Elemental abundance patterns of solar twins for the elements in the range 6 $\leq$$Z$$\leq$ 66 with respect to Fe for three groups classified by their ages. The mean ages are indicated. The observed abundances of solar twins are from \citet{Bedell_18}. Here, we remove the element Pr ($Z$=59) from our study since all stars exhibit a large deviation from the solar Pr/Fe ratio. For reference, offsets of 0.05 dex from the solar pattern (i.e., [X/Fe]=0) are shown by the dotted lines. In the top panel, the individual elemental abundance ratios measured thus far for bulge stars with [Fe/H]$\approx$0 are represented by the light-green stripes. The observed abundance data for Galactic bulge stars are from \citet[][for O, Mg, Ca, Mn, and Ni]{Schultheis_17}, \citet[][for Na, Al, Sc, Cr, Co, Cu, Zr, La, Ce, and Nd]{Duong_19}, \citet[][for Si]{Gonzalez_11}, \citet[][for S]{Lucertini_21}, \citet[][for Ti, Zn, Y, and Ba]{Bensby_17}, \citet[][for V]{Lomaeva_19}, and \citet[][for Eu]{Forsberg_19}. Regarding [C/Fe], we assign the observed range exhibited by three stars at [Fe/H]$\sim$0.2$-$0.4 \citep[the ratios could theoretically be higher at solar metallicity;][]{Romano_20}.
}
\end{figure}

Finally, we obtain five groups and show the elemental abundance patterns for the three of them according to age, i.e., $\langle {\rm age} \rangle$=8.7, 5.9, and 0.7 Gyr, in Figure 1. We clearly see the difference among the three patterns. Note that the other two groups with $\langle {\rm age} \rangle$=7.3 and 3.4 Gyr exhibit the intermediate patterns among the three in Figure 1. Before discussing Galactic chemical evolution based on these results, we elaborate on two findings. 

\subsection{Bulge migrators and the Sun's birthplace}

The first finding is that the abundance pattern of the oldest group is quite similar to that of Galactic bulge stars with [Fe/H]$\approx$0. In the top panel of Figure 1, the elemental abundances of the Galactic bulge around [Fe/H]=0 measured by several authors are superimposed over those of solar twins with $\langle {\rm age} \rangle$=8.7 Gyr. This good agreement between the two abundance patterns suggests that the oldest solar twins would migrate from inside of the bulge to the solar vicinity. The dynamic process of escaping from the Galactic bulge potential could be induced by perturbation of the bar \citep[][see also Di Matteo et al. 2014]{Raboud_98}. It should be noted that the fraction of local solar twins belonging to the oldest group is approximately 10\% (8/79) and thus is not negligible. This implies that migrators from the bulge would be an important population among nearby stars, particularly for super metal-rich stars.

Second, we find that the solar abundance pattern is identical to those of solar twins which range in age from 4.5 to 7.9 Gyr with $\langle {\rm age} \rangle$=5.9 Gyr and thus are typically older than the Sun (4.56 Gyr). This revealed association of the Sun with older solar twins suggests a common birthplace, specifically the region closer to the Galactic center. In this common region, interstellar matter was enriched to a mean metallicity [Fe/H] $\approx$ 0 approximately 6 Gyr ago, not 4.6 Gyr ago, with some variation in [Fe/H] over time. Considering the age of Galactic thin disk of $\sim$8$-$9 Gyr \citep[e.g.,][]{Fantin_19}, [Fe/H] reached the solar value within a short timescale of $\sim$2$-$3 Gyr at the Sun's birthplace. The implied metal-rich environment at the Sun's birthplace and formation time is supported by measured silicon isotopic ratios, i.e., $^{29}$Si/$^{28}$Si and $^{30}$Si/$^{28}$Si, in presolar silicon carbide grains \citep{Clayton_97, Tsujimoto_20}.

\subsection{Galactic chemical evolution across the disk}

\begin{figure}[t]
	\vspace{0.2cm}
    \hspace{0.3cm}
	\includegraphics[width=0.87\columnwidth]{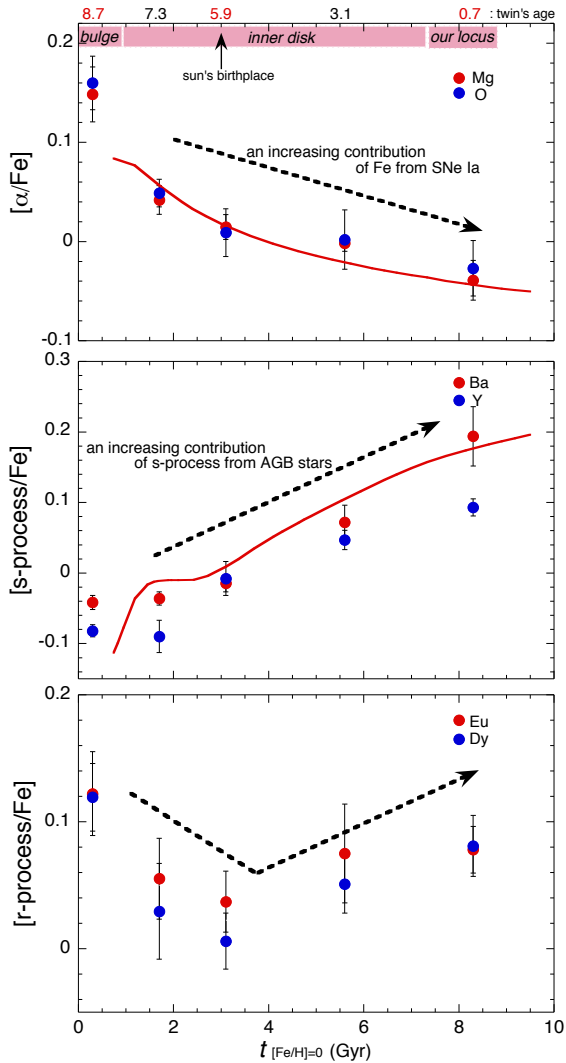}
	\vspace{-0.1cm}
\caption{{\it Top panel}: The correlation of the [$\alpha$/Fe] ratio for five groups of solar twins classified by mean stellar age relative to the elapsed time until the formation of individual twin groups, defined as $t_{\rm [Fe/H]=0}$, which is estimated from $\Delta t$ between the age of the disk (= 9 Gyr) and each mean age. The observed ratios are indicated by the filled circles for two elements (O and Mg), while the theoretical correlation of [Mg/Fe] predicted by the models for Galactic chemical evolution is shown by the red solid curve. This decreasing trend is interpreted as an outcome of an increasing contribution of Fe from SNe Ia into interstellar matter with increasing $t_{\rm [Fe/H]=0}$. {\it Middle panel}: The same as in the top panel but for [$s$-process/Fe] ($s$-process = Y and Ba). The predicted curve is for [Ba/Fe]. The increasing trend is attributed to a gradual accumulation of $s$-process elements in interstellar matter released from AGB stars in accordance with a longer elapsed time. {\it Bottom panel}: The correlation for [$r$-process/Fe]. The [$r$-process/Fe] ratio decreases with increasing $t_{\rm [Fe/H]=0}$ and upturns at approximately $t_{\rm [Fe/H]=0}$= 4 Gyr. These two discrete trends can be understood in the context of the properties of the $r$-process production site.
}
\end{figure}

Next, we argue how the variation in elemental abundance patterns in accordance with the mean age of solar twins can be understood within the framework of Galactic chemical evolution. A key hypothesis is that the abundance pattern of the older twin group is an end result of faster chemical enrichment at smaller $R_{\rm GC}$. By taking a close look at the transitions of [$\alpha$/Fe] and [$s$-process/Fe] with age, we show that a fundamental change in patterns is explained based on this hypothesis. First, we convert stellar ages to elapsed time, $t_{\rm [Fe/H]=0}$, for chemical enrichment up to [Fe/H]=0 from the start of star formation, assuming a disk age of 9 Gyr. Note that the oldest group with $\langle {\rm age} \rangle$=8.7 Gyr is considered to originate in the bulge, which is older than the disk, and that $t_{\rm [Fe/H]=0}$ for this group should, in practice, be larger than 0.3 Gyr. The top panel of Figure 2 shows the [$\alpha$/Fe] ratio as a function of $t_{\rm [Fe/H]=0}$ for five groups classified by mean stellar age (filled circles). We see a decreasing trend of [$\alpha$/Fe] with increasing $t_{\rm [Fe/H]=0}$. This is naturally interpreted to be an outcome of the larger amount of Fe released from SNe Ia into interstellar matter owing to more SNe Ia with longer delay times. In addition, the case for [$s$-process/Fe] is shown in the middle panel, which conversely exhibits an increasing trend. This feature is also reasonable if we consider a contribution of $s$-process elements from low- and intermediate-mass AGB stars which overwhelms the increase in Fe ejected from delayed SNe Ia. 

To validate the above arguments, we perform calculations of the Galactic chemical evolution. Models for the chemical evolution of the disk within the solar orbit (i.e., $R_{\rm GC}$ \ltsim 8 kpc) and of the bulge are designed to realize more efficient chemical enrichment as $R_{\rm GC}$ decreases, leading to a faster increase in metallicity and higher metallicity. We model each one for different $R_{\rm GC}$ by changing the timescales of star formation ($\tau_{\rm SF}$) and the supply of gas from the halo ($\tau_{\rm in}$). Here, star formation is assumed to be proportional to 1/$\tau_{\rm SF}$, and $\tau_{\rm in}$ is introduced in exponential form. The values of ($\tau_{\rm SF}$, $\tau_{\rm in}$) in units of Gyr are assumed to vary from (0.25, 0.3) for the bulge to (3.3, 5) for the solar vicinity. For the DTD of SNe Ia, we assume DTD $\propto t_{\rm delay}^{-1}$, with a range of $0.1\leq t_{\rm delay}\leq10$ Gyr \citep{Maoz_14}. We adopt the AGB yield for Ba which is empirically deduced from the abundances of Ba stars \citep{Tsujimoto_12}. The validity of this approach is guaranteed by agreement with the theoretical stellar yields \citep[][see also Prantzos et al.~2018 and Kobayashi et al.~2020]{Cseh_18}. Then, we calculate the evolutions of [Mg/Fe] and [Ba/Fe] and depict the results as red solid curves in the top and middle panels. Our results agree with the correlation of [Mg/Fe] and [Ba/Fe] with $t_{\rm [Fe/H]=0}$, that is, the mean age for solar twins (indicated at the top). This agreement confirms that the age-tagged abundances of solar twins are understood within the scheme of Galactic chemical evolution that combines the inside-out scenario with the galactic dynamics of radial migration.

\section{Galactic $r$-process enrichment}

\begin{figure*}[t]
	\hspace{2cm}
	\includegraphics[width=1.6\columnwidth]{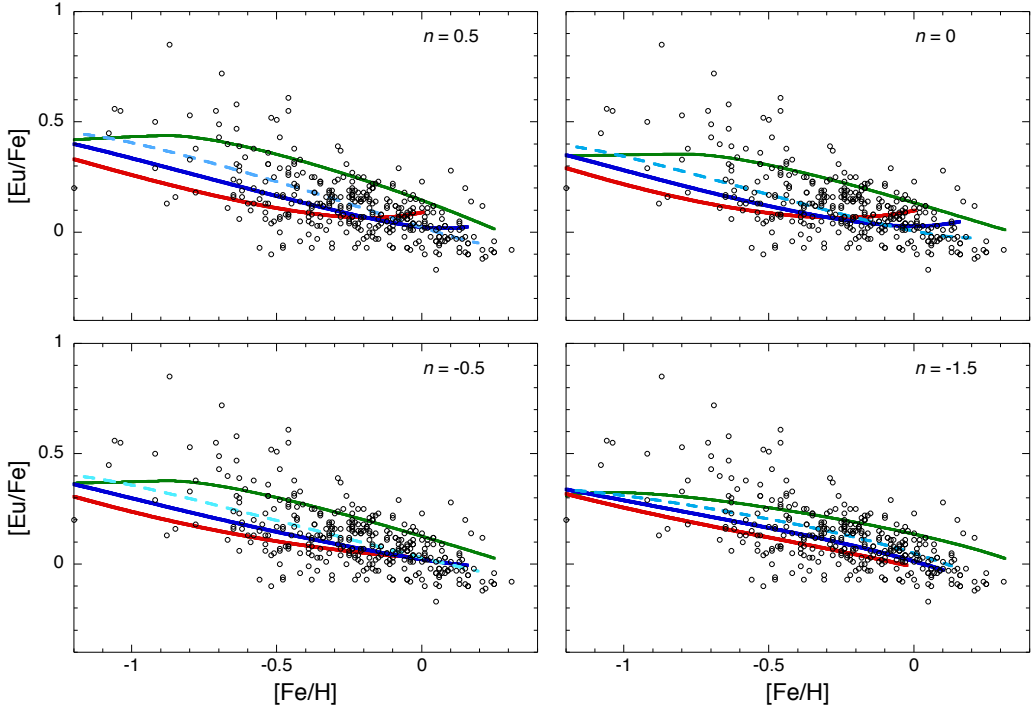}
	\vspace{0.0cm}
\caption{Evolutionary tracks of [Eu/Fe] predicted by models with different slopes ($n$=0.5, 0, -0.5, and -1.5) of the DTD for NSMs  compared with the observed data of nearby thin disk stars. The former three models include the contribution of $r$-process elements from rare CCSNe while NSMs are the sole site of the $r$-process in the model with $n$=-1.5. In each model, four tracks calculated at different chemical enrichment speeds (green, light blue, blue, and red, in order from fastest to slowest) are shown.}
\end{figure*}

In this context, the [$r$-process/Fe]-$t_{\rm [Fe/H]=0}$ correlation must be highlighted in terms of its direct association with two properties of $r$-process elements, i.e., their production sites and release timescales, both of which remain enigmatic. Here, we show the correlation of [$r$-process/Fe] with $t_{\rm [Fe/H]=0}$ for two elements, Eu and Dy{\footnote[1]{The Th abundance is measured for nearly 70\% of solar twins \citep{Botelho_19}. However, because of a large scatter in [Th/Fe] for each group arising from a large measurement error ($>$ 0.1 dex), we exclude Th from our discussion.}, in the bottom panel of Figure 2. Note that the percentages of each element produced by the $r$-process in terms of the solar abundance are 95\% and 85\% \citep{Prantzos_20}. Intriguingly, a peculiar feature$-$a decreasing trend followed by an upturn$-$is identified. These two discrete trends imply the mixture of two $r$-process sites with different, i.e., short and long, delay times for the product release, which can be associated with specific CCSNe and NSMs, respectively. 

Based on this consideration, we calculate the chemical evolution of Eu/Fe by using the models constructed for that of Mg/Fe and Ba/Fe, which are described in section 2.2. The key requirement for the models is to reproduce an upturn of [Eu/Fe] for $t_{\rm [Fe/H]=0}$ \gtsim 4 Gyr; this corresponding feature is largely determined by the DTD for NSMs. Conventionally, the DTD form for NSMs is parameterized by a power law ($\propto t_{\rm delay}^n$), with the power index being obtained from the DTD for short gamma-ray bursts (SGRBs). Although the index $n$ is poorly constrained, mainly owing to the difficulty in characterizing the SGRB host galaxy with a limited number of samples, studies to date are in rough agreement that $n$\gtsim$-1$ \citep[e.g.,][]{Zheng_07, Fong_13, Paterson_20}. In this study, we consider three cases of the DTD, i.e., $n$= -0.5, 0, and 0.5, with longer delay times chosen to reproduce the upturn feature, with a range of 0.03 $\leq t_{\rm delay} \leq 10$ Gyr. Together with these DTDs, we assume that NSMs occur at a rate of one per 300 CCSNe and that the ejected mass of Eu from each event is $2\times10^{-5}$\ms \citep{Tsujimoto_20}. In addition to NSMs, we assign specific CCSNe to a separate production site of $r$-process elements. For the production rate and yield, we adopt the results of magnetorotational SNe deduced from chemical evolution \citep{Tsujimoto_15} and nucleosynthesis calculations \citep{Nishimura_15}; Eu ejection is assumed to occur at a rate of one per 150 CCSNe,  with an ejected Eu mass of $1.2\times10^{-5}$\ms from each SN. In addition to these hybrid models involving two $r$-process sites, we  consider the case of a single $r$-process site of NSMs. For this model, we adopt a DTD with a steep slope of $-1.5$, which is constrained by the Galactic chemical evolution \citep{Hotokezaka_18, Cote_19}.

\begin{figure}[t]
	\vspace{0.3cm}
%    \hspace{1.6cm}
	\includegraphics[width=\columnwidth]{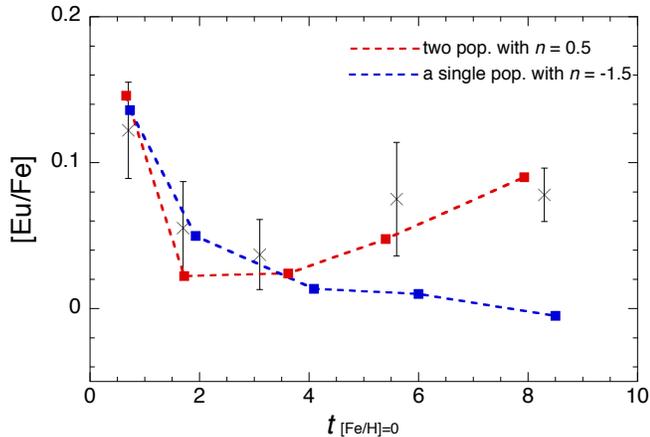}
%	\vspace{5cm}
\caption{The theoretical correlation of [Eu/Fe] predicted by Galactic chemical evolution models with $t_{\rm [Fe/H]=0}$ for the two models (dashed curves) compared with that for five groups of solar twins classified by the mean stellar age (crosses).}
\end{figure}

We first show four evolutionary paths of [Eu/Fe] against [Fe/H] calculated by individual models in Figure 3. Each path is a result of different speeds of chemical enrichment, corresponding to $t_{\rm [Fe/H]=0}$ = 0.3, 1.7, 3.1, and 8.3 Gyr. These model results are compared with the observed data of nearby thin disk stars. This comparison demonstrates that the four [Eu/Fe] curves in all models broadly overlap the domain of the observed data, indicating that all models pass the first test, i.e., agreement with the renewed view of Galaxy formation, in which stars in the solar vicinity represent a mixture of different stars born at various $R_{\rm GC}$. On the other hand, a close look at the trajectory of [Eu/Fe] at a late stage clarifies a difference among the four models; the models with $n$=0.5 and 0 predict a switch from a decreasing [Eu/Fe] trend to a slightly increasing one for the cases (red and blue curves) in which a relatively slow chemical enrichment proceeds, while the other two models predict a continious decreasing trend of [Eu/Fe] for all cases. Indeed, the former two model results are compatible with the correlation of [Eu/Fe] with $t_{\rm [Fe/H]=0}$ shown in the bottom panel of Figure 2.

This result is clearly demonstrated in Figure 4, which shows the sequence of [Eu/Fe] values at [Fe/H]=0 as a function of the elapsed time reaching  [Fe/H]=0 for the two models with $n$=0.5 and -1.5. We conclude that the joint $r$-process production of specific CCSNe and NSMs with a DTD comprising longer delay times ($n$\gtsim 0) explains the variation in $r$-process abundance among solar twins classified by their ages. Our obtained results agree with those of a previous study, which suggested $n$=0.6 as the most likely value within the allowable range of $-0.5< n <2.6$ \citep{Nakar_06}. 

\section{Summary}

The abundance patterns of solar twins are utilized to infer Galactic chemical evolution across the disk under the hypothesis that (i) their widely spanning ages are an outcome of radial migration from the inner region, including the bulge, and (ii) older solar twins were born closer to the Galactic center, where chemical enrichment occurs more quickly. This paradigm explains the variation in solar twin abundance patterns relative to their ages and provides new insight into the features of $r$-process production sites. We find that two distinct $r$-process sites should exist to accelerate $r$-process enrichment in both early and late phases of Galactic evolution. 

A population of $r$-process progenitors with long delay times for ejection is identified as NSMs whose power-law DTDs have a slope of $n$\gtsim 0. On the other hand, an $r$-process site with short delay times is likely to be associated with specific and rare CCSNe; however, this suggests the alternative possibility of the presence of a population ending promptly with NSMs \citep[e.g.,][]{Beniamini_19, Wanajo_21}. The CCSNe or NSMs for the short-delayed $r$-process site could be validated by future works, including an exploration of high-redshift SGRBs.

\acknowledgements

This work was performed in part at Aspen Center for Physics, which is supported by National Science Foundation grant PHY-1607611. The author thanks J. Baba for fruitful discussions on the stellar dynamics in relation to the Galactic bar. This work was supported by JSPS KAKENHI Grant Numbers 18H01258 and 19H05811.


\begin{thebibliography}{}
\bibitem[Baba et al.(2013)]{Baba_13}
Baba, J., Saitoh, T. R., \& Wada, K. 2013, \apj, 763, 46
\bibitem[Bedell et al.(2018)]{Bedell_18}
Bedell, M., Bean, J., Mel\'{e}ndez, J., et al. 2018, \apj, 865, 68
\bibitem[Beniamini \& Piran(2019)]{Beniamini_19}
Beniamini, P., \& Piran, T. 2019, MNRAS, 487, 4847
\bibitem[Bensby et al.(2017)]{Bensby_17}
Bensby, T., Feltzing, S., Gould, A., et al. 2017, A\&A, 605, A89
\bibitem[Botelho et al.(2019)]{Botelho_19}
Botelho, R. B., Milone, A. de C., Mel\'{e}ndez, J., et al. MNRAS, 482, 1690
\bibitem[Cavallo et al.(2021)]{Cavallo_21}
Cavallo, L., Cescutti, G., \& Matteucci, F. 2021, MNRAS, 503, 1
\bibitem[Chiappini et al.(2001)]{Chiappini_01}
Chiappini, C., Matteucci, F., \& Romano, D. 2001, \apj, 554, 1044
\bibitem[Clayton(1997)]{Clayton_97}
Clayton, D. D. 1997, \apj, 484, L67
\bibitem[C\^{o}t\'{e} et al.(2019)]{Cote_19}  
C\^{o}t\'{e}, B., Eichler, M., Arcones, A., et al. 2019, \apj, 875, 106
\bibitem[Cowperthwaite et al.(2017)]{Cowperthwaite_17}
Cowperthwaite, P. S., Berger, E., Villar, V. A., et al. 2017, \apjl, 848, L17
\bibitem[Cseh et al.(2018)]{Cseh_18}
Cseh, B., Lugaro, M., D'Orazi, V. et al., 2018, A\&A, 620, A146
\bibitem[Di Matteo et al.(2014)]{Di Matteo_14}
Di Matteo, P., Haywood, M., G\'{o}mez, A., et al. 2014, 567, A122
\bibitem[Dominik et al.(2012)]{Dominik_12}
Dominik, M., Belczynski, K., Fryer, C., et al. 2012, \apj, 759, 52
\bibitem[Duong et al.(2019)]{Duong_19}
Duong, L., Asplund, M., Nataf, D. M., Freeman, K. C., \& Ness, M. 2019, MNRAS, 486, 5349
\bibitem[Fantin et al.(2019)]{Fantin_19}
Fantin, N. J., C\^{o}t\'{e}, P., McConnachie, A. W., et al. 2019, \apj, 887, 148
\bibitem[Fong et al.(2017)]{Fong_17}
Fong, W., Berger, E., \& Blanchard, P. K., et al. 2017, \apjl, 848, L23
\bibitem[Fong \& Berger(2013)]{Fong_13}
Fong, W., \& Berger, E. 2013, \apj, 776, 18
\bibitem[Forsberg et al.(2019)]{Forsberg_19}
Forsberg, R., J\"{o}nsson, H., Ryde, N., \& Matteucci, F. 2019, A\&A, 631, A113
\bibitem[Gonzalez et al.(2011)]{Gonzalez_11}
Gonzalez, O. A., Rejkuba, M., Zoccali, M., et al. 2011, A\&A, 530, A54
\bibitem[Grand et al.(2012)]{Grand_12}
Grand, R. J. J., Kawata, D., \& Cropper, M. 2012, MNRAS, 421, 1529
\bibitem[Greggio et al.(2021)]{Greggio_21}
Greggio, L., Simonetti, P., \& Matteucci, F. 2021, MNRAS, 500, 1755
\bibitem[Hotokezaka et al.(2018)]{Hotokezaka_18}
Hotokezaka, K., Beniamini, P., \& Piran, T. 2018, Int. J. Mod. Phys., D 27, 1842005
\bibitem[Im et al.(2017)]{Im_17}
Im, M., Yoon, Y., Lee, S.-K. J., et al. 2017, \apjl, 849, L16
\bibitem[Ji et al.(2019)]{Ji_19}
Ji, A. P., Drout, M. R., \& Hansen, T. T. 2019, \apj, 882, 40
\bibitem[Kobayashi et al.(2020)]{Kobayashi_20}
Kobayashi, C., Amanda, K., \& Lugaro, M. 2020, \apj, 900, 179
\bibitem[Lomaeva et al.(2019)]{Lomaeva_19}
Lomaeva, M., J\"{o}nsson, H., Ryde, N., Schultheis, M., \& Thorsbro, B. 2019, A\&A, 625, A141
\bibitem[Lucertini et al.(2021)]{Lucertini_21}
Lucertini, F., Monaco, L., Caffau, E., Bonifacio, P., \& Mucciarelli, A. 2021, A\&A, in press
\bibitem[Maoz et al.(2014)]{Maoz_14}
Maoz, D., Mannucci, F., \& Nelemans, G. 2014, ARA\&A, 452, 107
\bibitem[Nakar et al.(2006)]{Nakar_06}
Nakar, E., Gal-Yam, A., \& Fox, D. B. 2006, \apj, 650, 281
\bibitem[Nishimura et al.(2015)]{Nishimura_15}
Nishimura, N., Takiwaki, T., \& Thielemann, F.-K. 2015, \apj, 810, 109
\bibitem[Paterson et al.(2020)]{Paterson_20}
Paterson, K., Fong, W., Nugent, A. et al. 2020, \apjl, 898, L32
\bibitem[Pian et al.(2017)]{Pian_17}
Pian, E., D'Avanzo, P., Benetti, S., et al. 2017, Nature, 551, 67
\bibitem[Prantzos et al.(2020)]{Prantzos_20}
Prantzos, N., Abia, C., Cristallo, S., Limongi, M., \& Chieffi, A. 2020, MNRAS, 491, 1832 
\bibitem[Prantzos et al.(2018)]{Prantzos_18}
Prantzos, N., Abia, C., Limongi, M., \& Chieffi, A. 2018, MNRAS, 476, 3432
\bibitem[Raboud et al.(1998)]{Raboud_98}
Raboud, D., Grenon, L., Martinet, L., Fux, R., \& Udry, S. 1998, 335, L61
\bibitem[Roederer(2013)]{Roederer_13}
Roederer, I. U. 2013, \apj, 145, 26
\bibitem[Romano et al.(2020)]{Romano_20}
Romano, D., Franchini, M., Grisoni, V., Spitoni, E., Matteucci, F., \& Morossi, C. 2020, A\&A, 639, A37
\bibitem[Ro\u{s}kar et al.(2008)]{Roskar_08}
Ro\u{s}kar, R., Debattista, V. P., Quinn, T. R., Stinson, G. S., \& Wadsley, J. 2008, \apjl, 684, L79
\bibitem[Schultheis et al.(2017)]{Schultheis_17}
Schultheis, M., Rojas-Arriagada, A., Garc\'{i}a P\'{e}rez, A. E., et al. 2017, A\&A, 600, A14
\bibitem[Sellwood \& Binney(2002)]{Sellwood_02}
Sellwood, J. A., \& Binney, J. 2002, MNRAS 336, 785
\bibitem[Siegel et al.(2019)]{Siegel_19}
Siegel, D. M., Barnes, J., \& Metzger, B. D. 2019, Nature, 569, 241
\bibitem[Smartt et al.(2017)]{Smartt_17}
Smartt, S.~J., Chen, T.-W., Jerkstrand, A., et al. 2017, Nature, 551, 75
\bibitem[Spina et al.(2018)]{Spina_18}
Spina, L., Mel\'{e}ndez, J., Karakas, A., et al. 2018, MNRAS, 474, 2580
\bibitem[Tsujimoto \& Baba(2020)]{TsujimotoBa_20}
Tsujimoto, T., \& Baba, J. 2020, \apj, 904, 137
\bibitem[Tsujimoto et al.(2020)]{Tsujimoto_20}
Tsujimoto, T., Nishimura, N., \& Kyutoku, K. 2020, \apj, 889, 119
\bibitem[Tsujimoto \& Nishimura(2015)]{Tsujimoto_15}
Tsujimoto, T., \& Nishimura, N. 2015, \apjl, 811, L10
\bibitem[Tsujimoto \& Bekki(2012)]{Tsujimoto_12}
Tsujimoto, T., \& Bekki, K. 2012, \apj, 747, 125
\bibitem[Wanajo et al.(2021)]{Wanajo_21}
Wanajo, S., Hirai, Y., \& Prantzos, N. 2021, MNRAS, 505, 5862
\bibitem[Winteler et al.(2012)]{Winteler_12}
Winteler, C., K\"{a}ppeli, R., Perego, A., et al. 2012, \apj, 750, L22
\bibitem[Zheng \& Ramirez-Ruiz(2007)]{Zheng_07}
Zheng, Z., \& Ramirez-Ruiz, E. 2007, \apj, 665, 1220
\end{thebibliography}
\end{document}